\documentclass[%
 reprint,
superscriptaddress,
showkeys,
 amsmath,amssymb,
 aps,
pre, 
longbibliography,
onecolumn
]{revtex4-2}

\usepackage{graphicx}
\usepackage[caption=false]{subfig} 

\usepackage{dcolumn}
\usepackage{bm}

\usepackage[a4paper, total={6.3in, 9in}]{geometry}

\newcommand{\ee}{\end{equation}} 
\newcommand{\be}{\begin{equation}}

\usepackage[mathscr]{euscript}  
\usepackage{xcolor}  

\usepackage{tcolorbox}
\usepackage{comment}
\usepackage{float}

\makeatletter
\newsavebox{\@brx}
\newcommand{\llangle}[1][]{\savebox{\@brx}{\(\m@th{#1\langle}\)}%
  \mathopen{\copy\@brx\kern-0.5\wd\@brx\usebox{\@brx}}}
\newcommand{\rrangle}[1][]{\savebox{\@brx}{\(\m@th{#1\rangle}\)}%
  \mathclose{\copy\@brx\kern-0.5\wd\@brx\usebox{\@brx}}}
\makeatother

\newcommand{\rmd}{{\rm d}}

\begin{document}

\preprint{ApS/123-QED}

\title{Partial stochastic resetting with refractory periods}

\author{Kristian St\o{}levik Olsen}
\thanks{kristian.olsen@hhu.de}

\affiliation{Institut für Theoretische Physik II - Weiche Materie, Heinrich-Heine-Universität Düsseldorf, D-40225 Düsseldorf, Germany}

\author{Hartmut L\"{o}wen}
\affiliation{Institut für Theoretische Physik II - Weiche Materie, Heinrich-Heine-Universität Düsseldorf, D-40225 Düsseldorf, Germany}

\begin{abstract}
The effect of refractory periods in partial resetting processes is studied.
Under Poissonian partial resets, a state variable jumps to a value closer to the origin by a fixed fraction at constant rate, $x\to a x$.  Following each reset, a stationary refractory period of arbitrary duration takes place. We derive an exact closed-form expression for the propagator in Fourier-Laplace space. For diffusive processes, we use the propagator to derive exact expressions for time dependent moments {\color{black} of $x$} at all orders. At late times the system reaches a non-equilibrium steady state which takes the form of a mixture distribution that splits the system into two subpopulations; trajectories that at any given time in the stationary regime find themselves in the freely evolving phase, and those that are in the refractory phase. In contrast to conventional resetting, partial resets give rise to non-trivial steady states even for the refractory subpopulation. Moments and cumulants {\color{black} associated with} the steady state {\color{black}density} are studied, and we show that a universal optimum for the kurtosis {\color{black}can be} found as a function of mean refractory time, determined solely by the strength of the resetting and the mean inter-reset time. The presented results could be of relevance to growth-collapse processes with periods of inactivity following a collapse.
\end{abstract}

\pacs{Valid pACS appear here} 
\maketitle

\section{Introduction}

Systems that exhibit substantial growth are often also susceptible to decay and collapse \cite{bardi2017seneca}. Unbounded growth is not physical in systems with finite resources, and complex systems that do not possess mechanisms for mitigating exaggerated growth may become unstable and vulnerable to sudden decay. External stimuli, such as accidents or other extreme events, may also cause such sudden disruption. Examples are observed in a wide range of complex systems, such as population numbers under disasters \cite{hanson1981logistic,gripenberg1983stationary}, crashes on the stock market \cite{bouchaud1998langevin}, and stress release during earthquakes or other forms of material failure \cite{vere1988variance,zheng1991application}. Also in cell biology we can find examples of growth-collapse phenomena, for example in a cell that grows and suddenly divides into two or more daughter-cells, effectively forcing the size of the mother-cell to collapse to some fractional value \cite{marantan2016stochastic,tanouchi2017long,wang2010robust}. From the perspective of statistical physics, stochastic resetting offers a powerful framework well-suited for the study of such recurrent extreme events. 

Stochastic resetting has emerged as a new branch of non-equilibrium statistical physics, where intriguing and surprising phenomena are aplenty. Over the last decade, this field has attracted the fascination of the physics community, in part due to the availability of steady states that, while arbitrarily far from equilibrium, are exactly solvable and brings insights into non-equilibrium phenomena. Since the work of Evans and Majumdar a little over a decade ago \cite{evans2011diffusion,evans2011optimal}, many extensions and generalizations have been considered. A myriad of systems has been exposed to the effects of conventional resetting, whereby an observable is instantaneously reset to its initial value at Poissonian instances of time, such as Brownian motion in potentials \cite{Pal_PRE,ray2020diffusion,KSO2023}, resetting in underdamped systems \cite{gupta2019stochastic,singh2020random,Olsen_2024}, and in active matter models \cite{evans2018run,baouche2024active,santra2021brownian,kumar2020active}. In addition,  a vast range of resetting schemes have been considered beyond the conventional setup. This includes non-Poissonian inter-reset durations \cite{pal2016diffusion,shkilev2017continuous,eule2016non,nagar2016diffusion,radice2022diffusion}, and resetting in finite time implemented by some physical resetting mechanism \cite{gupta2020stochastic,mercado2020intermittent,santra2021brownian,mercado2022reducing,Gupta_2021_SR,xu2022stochastic,roberts2024ratchetmediated}. The non-equilibrium nature of resetting systems have also been elucidated by stochastic thermodynamics \cite{fuchs2016stochastic,mori2023entropy,olsen2023thermodynamic,Deepak2022_work,busiello2020entropy,pal2023thermodynamic,goerlich2023experimental,olsen2024thermodynamic}. For a review, see Ref. \cite{evans2020review}.

An extension of conventional resetting that is particularly well-suited for dealing with the sudden collapse of complex systems is \emph{partial resetting}. This framework is able to account for the fact that sudden decay and collapse is often not complete but only partial. Here an observable $x(t)$ {\color{black} prepared initially in $x(0) = 0$, for example a particle's position,} is allowed to evolve following an arbitrary stochastic equation of motion, before being reset at random times to $a x(t)$, with a \emph{resetting strength}. {\color{black}Since the process is only reset partially towards its initial state $x=0$ for $a\in (0,1)$, we consider only this range of values for $a$, although in principle other values could be considered, making the process a hybrid diffusion/jump process \cite{harbola2023stochastic,harbola2023stochastic2,olsen2024thermodynamic}.} Within the applied mathematics community, similar processes have been studied under the guise of Markovian growth-collapse models, where typically a deterministic growth is interrupted by instantaneous decays \cite{privault2022moments, boxma2006markovian, lopker2011hitting}. {\color{black} Other systems that share some formal similarity to partial resetting processes can be found for example in the dynamics of cells under division \cite{hall1989functional}, and particles undergoing inelastic collisions with a vibrating plate \cite{majumdar2007inelastic}.}
In the modern version of stochastic resetting, partial resets were first studied in the context of advection-diffusion processes \cite{pierce2022advection,tal2022diffusion}. Since then, several studies have surfaced, investigating aspects such as time-dependent propagators for general Markovian systems, thermodynamics, and first passage times \cite{di2023time, olsen2024thermodynamic,biroli2024resetting}.

\begin{figure*}
    \centering
    \includegraphics[width = \textwidth]{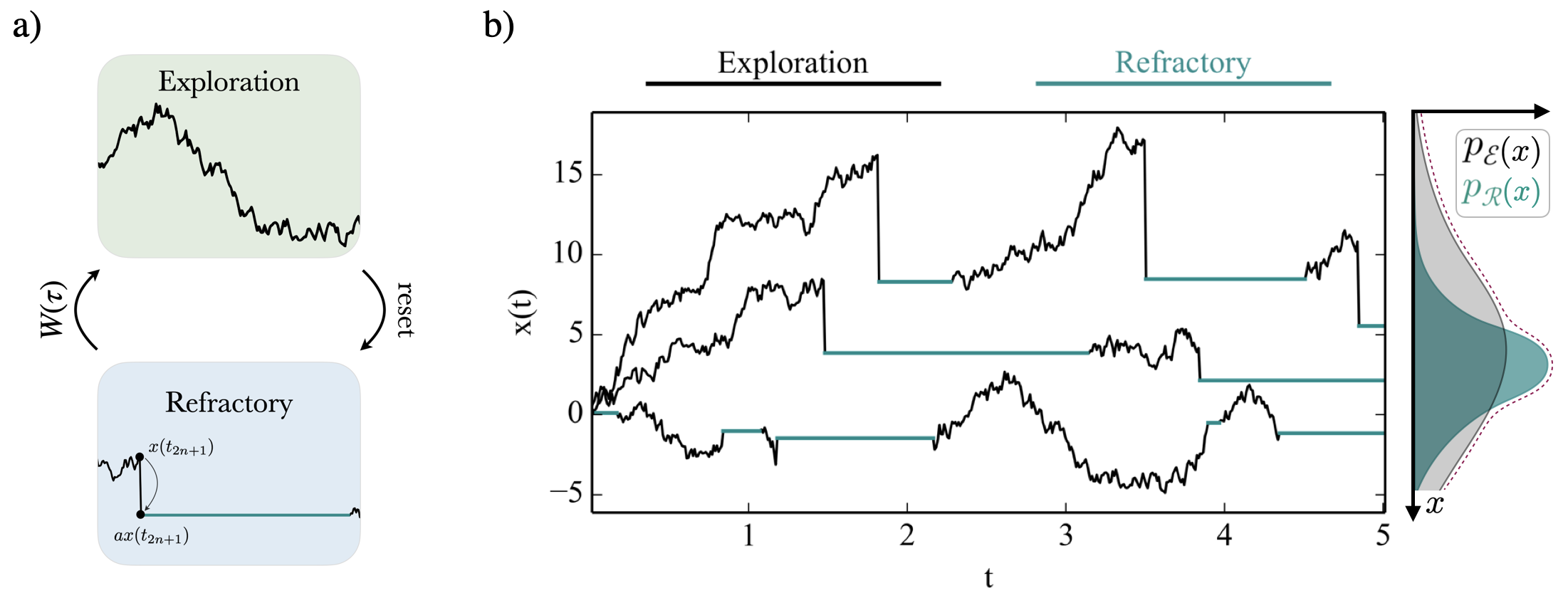}
    \caption{a) In the exploration state $\mathcal{E}$ the system resets at rate $r$ and enters the refractory state $\mathcal{R}$. After a time drawn from a density $W(\tau)$, a new exploration state in initiated. b) Sketch of typical {\color{black} evolution of the state variable $x(t)$}. Stochastic trajectories undergo partial resetting with strength $a=1/2$.  At any given time in the stationary regime, a constant fraction of trajectories in a large ensemble occupy the exploration and refractory states. This is reflected in the steady state, which can be decomposed into two subpopulations $\rho_\mathcal{S}(x)$, $\mathcal{S} = \mathcal{E},\mathcal{R}$, corresponding to particles in the exploration and refractory states respectively. }
    \label{fig:fig1}
\end{figure*}

Here, we study partial resetting under the additional effect of refractory periods. Refractory periods are phases where, following a reset, the state remains idle for a random duration $\tau$ drawn from a distribution $W(\tau)$ \cite{evans2018effects,maso2019stochastic}. After this idle period, a new exploration phase is initiated (see Fig.(\ref{fig:fig1})). This can be of interest for several reasons. { \color{black} First, one may easily imagine that in natural systems that undergo a sudden collapse, there is a period of inactivity following the collapse. Examples include populations whose growth may briefly be stifled by a collapse, stock market collapse where trader scepticism prevents immediate growth, or in the cell cycle where the DNA content does not immediately grow after mitosis.} Second, the mixed effect of partial resets and refractory times may give rise to intriguing steady state properties and be of interest from a theoretical perspective. Indeed, for conventional resetting (corresponding to $a = 0$) the presence of refractory periods has been shown to give rise to a steady state that is a weighted mixture of the steady state in the absence of refractory periods and a Dirac delta-function located at the resetting position \cite{evans2018effects,garcia2024stochastic_pub}.  
The mean refractory time $\langle \tau \rangle$ then defines a one-parameter family of steady states that interpolates between the refractory-free steady state  (at $\langle \tau \rangle=0$) and a Dirac delta function (as $\langle \tau \rangle\to \infty$). The appearance of a Dirac delta in the steady state is a simple consequence of the fact that after every reset, the process remains idle at the resetting position for some refractory time. However, in the partial resetting scenario the resetting position is never the same, and more complex steady state properties are to be expected. Furthermore, it is known that partial resetting gives rise to steady states that transition from non-Gaussian shapes at strong resetting, to Gaussian at weak resetting \cite{tal2022diffusion}. The inclusion of both refractory times and partial resets will enable us to see whether these transitions persist, and how the two effects act in conjunction.

This paper is organized as follows. Section \ref{sec:prop} derives an exact expression for the propagator in Fourier-Laplace space, and consider various limiting scenarios. Section \ref{sec:time} consider the case of a diffusion process in detail, where explicit expressions for time-dependent moments are calculated. Section \ref{sec:ness} studies the non-equilibrium steady state and its behaviour under various choices of refractory times and resetting strengths, before section \ref{sec:concl} offer a concluding discussion.

\section{The propagator}\label{sec:prop}

We consider a process that alternates between an exploring phase and a refractory phase following a reset. In the refractory phase, no evolution takes place. We denote the times at which the process switches from one phase to another $t_i$, such that exploration periods take place in $t\in (t_{2n-2},t_{2n-1})$ and refractory periods in $t\in (t_{2n-1},t_{2n})$. We denote the intervals
\begin{align}
    \mathcal{T}_n &= t_{2n-1}-t_{2n-2},\\
    \tau_n &= t_{2n}-t_{2n-1},
\end{align}
and denote their distributions by $\mathcal{T} \sim \psi(\mathcal{T})$ and $\tau \sim W(\tau)$ respectively. While in the exploration phase, the state evolves according to underlying (reset-free) propagator $p_0(x,t|x_0)$. At the end of an exploration phase,  the particle resets partially and transitions into a refractory state, where remains at this location for a duration drawn from $W(\tau)$ (see Fig.(\ref{fig:fig1})). We consider exploration phases with exponential durations $\psi(\mathcal{T}) = r e^{-r \mathcal{T}}$, and keep $W(\tau)$ arbitrary for the moment.

To proceed analytically, we make use of the renewal structure present in most resetting problems \cite{evans2020review}. Here it is convenient to use a first renewal equation, which in this instance takes the form \cite{evans2018effects}
\begin{align}
    p_r(x,t|x_0) &= e^{-rt}p_0(x,t|x_0)\nonumber\\
    & + r \int_0^t \rmd t_1 e^{-r t_1}\int_0^{t-t_1}\rmd\tau W(\tau)\int \rmd y  p_0(y,t_1|x_0) p_r(x,t-t_1-\tau|a y) \nonumber \\
    & + r\int_0^t \rmd t_1 e^{-r t_1} \int_{t-t_1}^\infty \rmd \tau W(\tau) \int \rmd y  p_0(y,t_1|x_0)\delta(x-ay).
\end{align}

Here the first term corresponds to trajectories where no resets take place up to time $t$, and the system evolves according to the underlying propagator. The probability that no reset takes place is simply $e^{- r t}$. The second term takes into account trajectories that at time $t$ are in the exploration phase. The system evolves from the initial state $x_0$ to a random position $y$ in time $t_1$. It then resets $y\to a y$, and remains at this new position for a time $\tau$. In the remaining time $t-t_1 -\tau$ the particle propagates to the final state $x$. The third term takes into account trajectories that at time $t$ end in the refractory phase. After an evolution from the initial state $x_0$ to a random position $y$ followed by subsequent partial reset $y\to a y$, the particle now remains in the refractory phase at least until time $t$, i.e. $\tau \in (t-t_1,\infty)$. These trajectories can only contribute to the propagator $p_r(x,t|x_0) $ if $x = a y$, which is the reason for the Dirac delta function in the above renewal equation. To proceed, we perform a Laplace transform, following Ref. \cite{evans2018effects}, and make use of the convolution theorem. This results in 
\begin{align}\label{eq:hard}
    \tilde p_r(x,s|x_0) &= \tilde p_0(x,s+r|x_0) + r \tilde W(s) \int \rmd y  \tilde p_0(y,s+r|x_0) \tilde p_r(x,s|a y) \nonumber \\
    &+ r\frac{1-\tilde W(s)}{a s} \tilde p_0(x/a,s+r|x_0).
\end{align}
A similar Laplace transform was performed in Ref. \cite{evans2018effects} for conventional resetting, and we refer the reader to this reference for further details.

To obtain a closed relation for $\tilde p_r(x,s|x_0)$ we need to deal with the remaining integral over the intermediate position $y$. The expression is almost in the form of a convolution, however, $\tilde p_r(x,s|a y) \neq \tilde p_r(x- a y,s|0)$ since partial resetting introduces a spatial heterogeneity that is not present in the underlying system without resetting. This is a simple consequence of the fact that the length of the resetting step, for fixed strength $a$, depends on the initial position of the particle. However, these issues may be circumvented by iterating Eq.\,(\ref{eq:hard}) multiple times, which leads to 

 {
 \color{black}
\begin{equation}\label{eq:fullprop}
         \tilde p_r(x,s|x_0) =   \mathcal{P}_\mathcal{E} (x,s|x_0) +   \mathcal{P}_\mathcal{R} (x,s|x_0),
\end{equation}
where we introduced
\begin{align}\label{eq:pe}
    \mathcal{P}_\mathcal{E} (x,s|x_0) &= \tilde p_0(x,s+r|x_0) +r \tilde W(s) \int \rmd y  \tilde p_0(y,s+r|x_0) \tilde p_0(x,s+r|a y) \nonumber\\
    & + (r\tilde W(s))^2 \int \rmd y \rmd y'  \tilde p_0(y,s+r|x_0) \tilde p_0(y',s+r|ay) \tilde p_0(x,s+r|a y') + ...
\end{align}
which incorporates all the contributions from trajectories that end in the exploration phase, and
\begin{align}\label{eq:pr}
    \mathcal{P}_\mathcal{R} (x,s|x_0) &=  r \frac{1-\tilde W(s)}{a s} \tilde p_0(x/a,s+r |x_0)  + r \frac{1-\tilde W(s)}{a s}  (r \tilde W(s)) \int \rmd y \tilde p_0(y,s+r |x_0)  \tilde p_0(x/a,s+r |ay)  \nonumber \\
    &+ r \frac{1-\tilde W(s)}{a s}  (r \tilde W(s))^2 \int \rmd y \rmd y' \tilde p_0(y,s+r |x_0) p_0(y',s+r |a y)  \tilde p_0(x/a,s+r |a y') +...
\end{align}
which incorporates all the contributions from trajectories that end in the refractory phase. Diagrammatically, and for the sake of easier readability, the propagator can be expressed diagrammatically as

\begin{figure}[h!]
    \centering
    \includegraphics[width = 12cm]{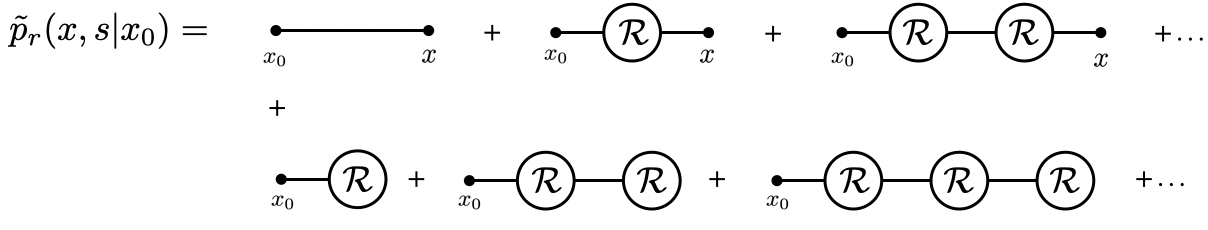}
    \label{fig:diag}
\end{figure}

\noindent where each solid line corresponds to a propagator of the underlying system with appropriate initial conditions and $\mathcal{R}$ denotes refractory periods.  Here, the first row corresponds to $\mathcal{P}_\mathcal{E} (x,s|x_0)$, and the second row of diagrams to $\mathcal{P}_\mathcal{R} (x,s|x_0)$. } Each \emph{complete} refractory period is accompanied by a factor of $r \tilde W(s)$, and trajectories ending in a refractory period has one additional factor of $r (1-\tilde W(s))/s$. For trajectories ending in a refractory phase, they arrive at some random location at the time of entering this last phase, and remain there. 

With Eq.\,(\ref{eq:fullprop})-(\ref{eq:pr}), the propagator is expressed entirely in terms of the underlying propagator, for which we are free to use spatial homogeneity to write $\tilde p_0(x,s|x_0) = \tilde p_0(x - x_0,s|0)$. All integrals are then converted to convolutions, and we can perform a Fourier transform and use the convolution theorem to obtain the solution
\begin{align}
        \hat{\tilde p}_r(k,s|x_0) &=\sum_{n=0}^\infty [r \tilde W (s)]^n  e^{- i a^n k x_0}\prod_{j=0}^n \hat{\tilde{p}}_0(a^j k, s+r |0) \nonumber \\
        & + r \frac{1-\tilde W(s)}{s}\sum_{n=1}^\infty [r \tilde W(s)]^{n-1} e^{- i a^n k x_0}\prod_{j=1}^n \hat{\tilde{p}}_0(a^j k, s+r |0). \label{eq:exact}
\end{align}
This is an exact result, valid for any distribution of refractory periods $W(\tau)$ and any underlying system with propagator $p_0(x,t|x_0)$. This result extends previous studies, which can be recovered by taking various limits. Some special cases worth highlighting are:

\begin{itemize}
    \item[i)]  \textit{Complete resetting $a = 0$}: When resetting is complete, the propagators in the products in Eq. (\ref{eq:exact}) will due to normalization satisfy $\hat{\tilde{p}}_0(a^j k, s+r |0) = (s+r)^{-1}$ for $j \geq 1$. Only the $j=0$ terms contributes to the propagator $\hat{\tilde p}_r(k,s|x_0)$. Furthermore, since the system is homogeneous, we set $x_0 =0$. We find 
\begin{align}
        \hat{\tilde p}_r(k,s|0) &=\hat{\tilde{p}}_0( k, s+r |0)  \sum_{n=0}^\infty [r \tilde W (s)]^n (s+r)^{-n}  +  \frac{1-\tilde W(s)}{s  \tilde W(s)}\sum_{j=1}^\infty [r \tilde W(s)]^n (s+r)^{-n} \\
        & = \frac{(s+r)\hat{\tilde{p}}_0( k, s+r |0)  + \frac{r}{s} [1-\tilde W(s)]}{s + r - r \tilde W(s)},
\end{align}
which is exactly the solution obtained in Refs. \cite{evans2018effects,garcia2024stochastic_pub} for this particular scenario.
\item[ii)] \textit{No refractory period $W(\tau) = \delta(\tau):$} when the refractory periods have vanishing duration, $\tilde W(s) = 1$. This immediately removes the last term in Eq.\,(\ref{eq:exact}), and we have
\begin{align}
        \hat{\tilde p}_r(k,s|x_0) &=\sum_{n=0}^\infty r^n  e^{- i a^n k x_0}\prod_{j=0}^n \hat{\tilde{p}}_0(a^j k, s+r |0).
\end{align}
This is the solution studied in Ref \cite{di2023time} for partial resetting without refractory periods.
\item [iii)] \textit{Weak resetting $a = 1$:} In the weak resetting limit, the propagator will now still deviate significantly from the underlying propagator due to the presence of refractory times. Indeed, letting $a=1$ in Eq.\,(\ref{eq:exact}) we find 
\begin{align}\label{eq:sng}
        \hat{\tilde p}_r(k,s|x_0) &= \frac{\hat{\tilde{p}}_0( k, s+r |x_0) }{1 - r \tilde W(s) \hat{\tilde{p}}_0( k, s+r |0) } \left( 1 + \frac{r}{s} [1-\tilde W(s)]  \right).
\end{align}
In Ref.  \cite{Olsen_2024} this propagator was derived for systems that undergo stop-and-go motion using a velocity resetting protocol. Since $a=1$, the resetting does not alter the particle's position, but nonetheless the particle enters into a stationary refractory phase which last for a random duration. 
\end{itemize}

\section{Time-evolution of moments for diffusion processes}\label{sec:time}
Before examining general properties of steady states under partial resets with refractory times, we consider in detail the time dependence of moments when the reset-free system is purely diffusive and $x_0 = 0$. At early times  when no resets have taken place, we expect that the moments behave as in the purely diffusive case, with $\langle x^{2\ell}\rangle \sim t^\ell$. At late times, we expect a steady state value to be reached. However, the transient behaviour connecting these regimes can be rather complex.

For diffusion, the underlying propagator takes the form 
\begin{equation}
    \hat{\tilde{p}}_0( k, s |0)  = \frac{1}{s + D k^2}.
\end{equation}
Using Eq.\,(\ref{eq:exact}) we can write the full resetting propagator as

\begin{align}
        \hat{\tilde p}_r(k,s|x_0) &=\sum_{n=0}^\infty \frac{[r \tilde W (s)]^n}{(s+r)^{n+1}}  \prod_{j=0}^n \frac{1}{1 + \frac{D a^{2j}k^2}{s+r}}  + r \frac{1-\tilde W(s)}{s}\sum_{n=1}^\infty \frac{[r \tilde W(s)]^{n-1}}{(s+r)^{n}} \prod_{j=1}^n \frac{1}{1 + \frac{D a^{2j}k^2}{s+r}}.
\end{align}
To proceed, we notice the appearance of the q-Pochhammer symbol
\begin{equation}
    (x;y)_{n+1} = \prod_{j=0}^{n} (1- x y^j),
\end{equation}
which leads to a propagator of the form
\begin{align}
        \hat{\tilde p}_r(k,s|x_0) &=\sum_{n=0}^\infty \frac{[r \tilde W (s)]^n}{(s+r)^{n+1}}  \left( \frac{D k^2}{s+r} ; a^2\right)_{n+1}^{-1} \nonumber \\
        & + r \frac{1-\tilde W(s)}{s}\sum_{n=1}^\infty \frac{[r \tilde W(s)]^{n-1}}{(s+r)^{n}}\left(1 + \frac{D k^2}{s+r}\right)  \left( \frac{D k^2}{s+r} ; a^2\right)_{n+1}^{-1}.
\end{align}
The q-Pochhammer symbol is a well-studied function and has a rich mathematical theory with connections to number theory, modular forms and the partition of integers \cite{hardy1979introduction}. Series representations, known as q-series, are well-established \cite{koekoek1996askey,gasper1995lecture}, including
\begin{equation}\label{eq:qpoc}
     (x;y)_{n+1}^{-1} =\sum_{\ell =0}^\infty {\ell + n \brack \ell}_y x^\ell ,
\end{equation}
where the square bracket denotes the Gaussian binomial coefficients. Using this series representation, we can write the propagator as

\begin{align}
     \hat{\tilde p}_r(k,s|x_0) &=  \frac{1}{s}  + \sum_{\ell=1}^\infty (-1)^\ell \left(  \sum_{n=0}^\infty {\ell + n \brack \ell}_{a^2} \frac{[r \tilde W(s)]^n}{(s+r)^{n+\ell +1}} d^\ell  \right) k^{2\ell} \nonumber \\
     &+\sum_{\ell = 1}^\infty  (-1)^\ell \left(  \sum_{n=1}^\infty {{\ell} + n \brack {\ell}}_{a^2} \frac{ a^{2\ell} (1-a^{2n}) }{1 - a^{2(\ell+n)}}    r \frac{1-\tilde W(s)}{s}  \frac{[r \tilde W(s)]^{n-1}}{(s+r)^{n+\ell}}    D^\ell \right)  k^{2\ell}. \label{eq:char}
\end{align}
For details of the derivation, see appendix \ref{sec:moment_time}.  This can be compared with the characteristic series
\begin{equation}
     \hat{\tilde p}_r(k,s|x_0) = \sum_{\ell = 0}^\infty \frac{(-1)^\ell}{(2\ell)!} \widetilde {\langle x^{2\ell} \rangle} k^{2\ell}
\end{equation}
to identify the (Laplace transformed) moments of order $2\ell$:
\begin{align}
    \widetilde {\langle x^{2\ell} \rangle}  &=  (2\ell)!  \sum_{n=0}^\infty {\ell + n \brack \ell}_{a^2} \frac{[r \tilde W(s)]^n}{(s+r)^{n+\ell +1}} D^\ell   \\
     &+ (2\ell)!   \sum_{n=1}^\infty {{\ell} + n \brack {\ell}}_{a^2} \frac{ a^{2\ell} (1-a^{2n}) }{1 - a^{2(\ell+n)}}    r \frac{1-\tilde W(s)}{s}  \frac{[r \tilde W(s)]^{n-1}}{(s+r)^{n+\ell}}    D^\ell   ,
\end{align}
which holds for $\ell \geq 1$.  To obtain the full time evolution of the moments in the general case, we must perform an inverse Laplace transform of the above coefficients
\begin{align}\label{eq:timemoms}
      {\langle x^{2\ell} \rangle} &=  (2\ell)!  \sum_{n=0}^\infty {\ell + n \brack \ell}_{a^2} \mathcal{L}^{-1}_t\left(\frac{[r \tilde W(s)]^n}{(s+r)^{n+\ell +1}} \right) D^\ell   \\
     &+ (2\ell)!   \sum_{n=1}^\infty {{\ell} + n \brack {\ell}}_{a^2} \frac{ a^{2\ell} (1-a^{2n}) }{1 - a^{2(\ell+n)}}    r \mathcal{L}^{-1}_t\left( \frac{1-\tilde W(s)}{s}  \frac{[r \tilde W(s)]^{n-1}}{(s+r)^{n+\ell}}  \right)  D^\ell  . \nonumber
\end{align}
This gives the exact moments and their time evolution for any $W(\tau)$, with an inverse Laplace transform to be calculated in each case. In the limit $r\to 0$ only the first term of the first sum survives, and we recover the diffusive moments $ {\langle x^{2\ell} \rangle} = (2 D t)^\ell (2\ell-1)!!$ as expected. Next we consider two choices for $W(\tau)$.\\

\noindent\textbf{i) No refractory times: } In the special case of no refractory times $\tilde W(s) = 1$ and we have

\begin{align}
    \langle x^{2\ell} \rangle &=  (2\ell)!  \sum_{n=0}^\infty {\ell + n \brack \ell}_{a^2} \mathcal{L}^{-1}_t\left(\frac{r^n}{(s+r)^{n+\ell +1}} \right) D^\ell  \\
     & =  (2\ell)!  \sum_{n=0}^\infty {\ell + n \brack \ell}_{a^2} \left(\frac{D}{r}\right)^\ell  \frac{ (r t) ^{n+\ell}}{(n+\ell)!}  e^{- r t}.
\end{align}
This gives the exact time-dependent moments, albeit in the form of an infinite sum. For $\ell = 1$, the sum for the second moment can be carried out explicitly, and we find 
\begin{equation}
  \langle x^{2} \rangle =  \frac{2 D \left(1 -e^{- \left(1-a^2\right) r t} \right)}{r \left(1-a^2\right) },
\end{equation}
which coincides with the particular case calculated in Ref. \cite{tal2022diffusion}.\\

\begin{figure}[t!]
    \centering
    \includegraphics[width = \textwidth]{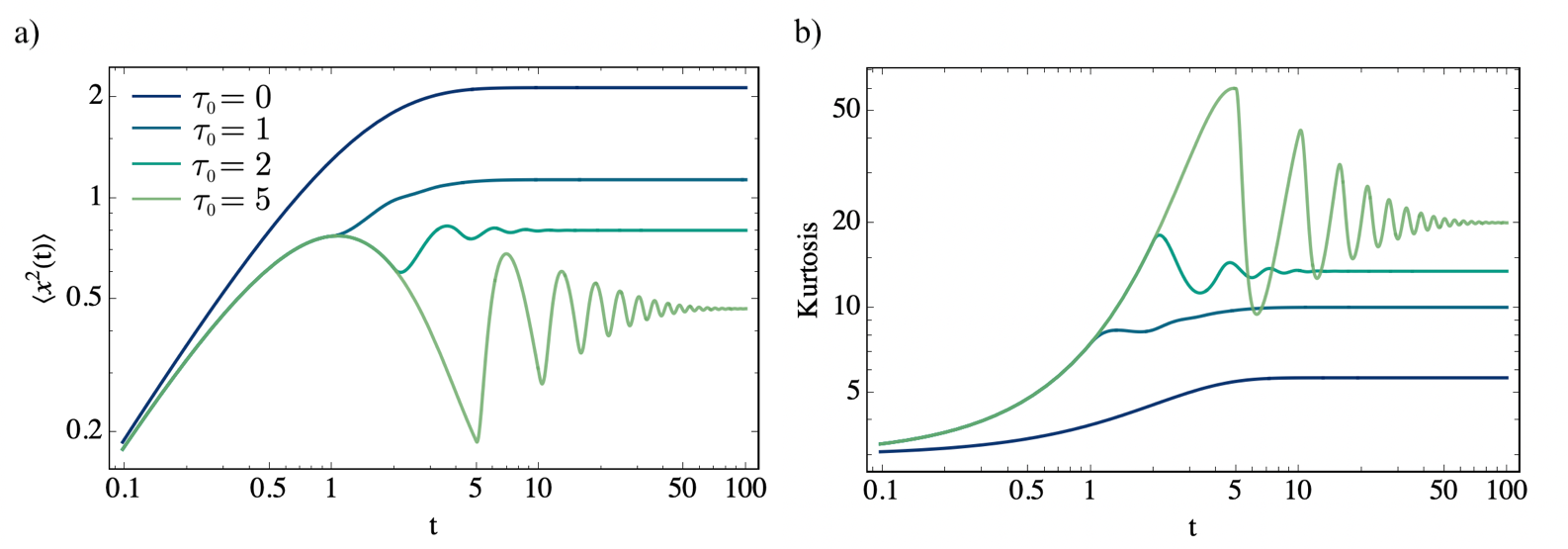}
    \caption{Analytic solution for the mean squared displacement (a) and kurtosis (b) for a diffusion process with sharp refractory durations $W(\tau) = \delta(\tau-\tau_0)$ and partial resetting with strength $a = 1/4$, obtained from Eq.\,(\ref{eq:timemoms}) and Eqs. (\ref{eq:invl1})\& (\ref{eq:invl2}).  Other parameters are set to $D = 1, r = 1$.}
    \label{fig:msd_oscil}
\end{figure}

\noindent\textbf{ii) Sharp refractory times: } For sharp refractory times with fixed duration $\tau_0$ we have $W(\tau) = \delta(\tau-\tau_0)$, and $\tilde W(s) = \exp(- s \tau_0)$. The inverse Laplace transforms needed in Eq.\,(\ref{eq:timemoms}) then can be inverted using numerical software, resulting in 
{\color{black}
\begin{align}
    \mathcal{L}^{-1}_t\left(\frac{[r \tilde W(s)]^n}{(s+r)^{n+\ell +1}} \right) &=  \theta \left(t-n \tau _0\right) \frac{r^n  \left(t-n \tau _0\right){}^{l+n} e^{-r \left(t-n \tau _0\right)}}{\Gamma
   (l+n+1)},  \label{eq:invl1}\\
     \mathcal{L}^{-1}_t\left( \frac{1-\tilde W(s)}{s}  \frac{[r \tilde W(s)]^{n-1}}{(s+r)^{n+\ell}}  \right)  &= -\theta(t- n\tau_0) \frac{r^{-l-1} \left[ \Gamma(n+\ell) - \Gamma(n+\ell, r (t - n \tau_0))\right]}{\Gamma[n+\ell]}  \label{eq:invl2}  \\
    & + \theta(t- n\tau_0+\tau_0) \frac{r^{-l-1} \left[ \Gamma(n+\ell) - \Gamma(n+\ell, r (t - n \tau_0+\tau_0))\right]}{\Gamma[n+\ell]}.\nonumber
\end{align}
}
where $\theta(x)$ is the Heaviside theta-function and $\Gamma(x,y)$ the incomplete gamma-function. Using Eq.\,(\ref{eq:timemoms}), we plot the mean squared displacement $\langle x^{2} \rangle$ and kurtosis $\langle x^{4} \rangle/[\langle x^{2} \rangle]^2$  in Fig.\,(\ref{fig:msd_oscil}) for various values of the refractory time $\tau_0$. The peaks observed in the mean squared displacement correspond approximately to times when the particle is at the end of an exploration phase and is about to reset, i.e. at times $t_n = (n-1) [\tau_0+1/r] + 1/r$ where $(n-1) [\tau_0+1/r]$ is the mean time of $n-1$ exploration and refractory phases, and the additional waiting time $1/r$ takes the particle to its position just before the $n$'th reset. The kurtosis and displays the same non-monotonic behaviour, although the peaks are somewhat shifted when compared to the case of the mean squared displacement.  At early times, the kurtosis takes the Gaussian value $3$, since no resetting has affected the diffusive dynamics at this point. We see that the steady state value of the kurtosis reached at late times is always greater than $3$, indicating that the steady states are leptokurtic. The properties of these non-equilibrium steady states are investigated further in the next section.


\section{Non-equilibrium steady states}\label{sec:ness}
Having studied dynamical properties of partial resets  with refractory times for the particular case of diffusion, we next turn to properties of non-equilibrium steady states for general systems. These steady states can be extracted from Eq.\,(\ref{eq:exact}) through application of the final value theorem
\begin{equation}\label{eq:finalval}
    \hat p_r^*(x) \equiv \lim_{s\to 0} s  \hat{\tilde p}_r(k,s|x_0).
\end{equation}
We first rewrite the expression for the propagator as
\begin{align}\label{eq:startingpoint}
        \hat{\tilde p}_r(k,s|x_0) &=\sum_{n=0}^\infty \left[\frac{r \tilde W (s)}{s+r}\right]^n \frac{1}{s+r} e^{- i a^n k x_0}\prod_{j=0}^n (s+r)\hat{\tilde{p}}_0(a^j k, s+r |0) \nonumber \\
        & + r \frac{1-\tilde W(s)}{s}\sum_{n=1}^\infty \left[\frac{r \tilde W (s)}{s+r}\right]^{n-1}  \frac{1}{s+r} e^{- i a^n k x_0}\prod_{j=1}^n (s+r)\hat{\tilde{p}}_0(a^j k, s+r |0).
\end{align}
This makes the factors in the products well-behaved, with $(s+r)\hat{\tilde{p}}_0(a^j k, s+r |0)$ simply approaching unity for large values of $j$. To proceed, we want to use the final value theorem, Eq.\,(\ref{eq:finalval}), and extract the coefficient in front of the $s^{-1}$ pole at small $s$-values in these expressions. However, taking a small-$s$ limit may not commute with the infinite sums in Eq.\,(\ref{eq:startingpoint}). For technical details of performing this limit, see appendix \ref{sec:nesscalc}. The non-equilibrium steady state is found to take the form
\begin{equation}\label{eq:stst}
    \hat{p}_r^*(k) =  \frac{1}{ 1+ r \langle \tau \rangle}  \prod_{j=0}^\infty r\hat{\tilde{p}}_0(a^j k, r |0)  +  \frac{r \langle \tau \rangle}{ 1+ r \langle \tau \rangle}  \prod_{j=1}^\infty r\hat{\tilde{p}}_0(a^j k, r |0).
\end{equation}
When there are no refractory periods $( \langle \tau \rangle=0)$, we recover the results of Ref.\,\cite{tal2022diffusion}. When the resetting is very strong $(a\to 0)$ we recover the results of conventional resetting with refractory times \cite{evans2018effects, garcia2024stochastic_pub}.

\begin{figure}
    \centering
    \includegraphics[width = 15cm]{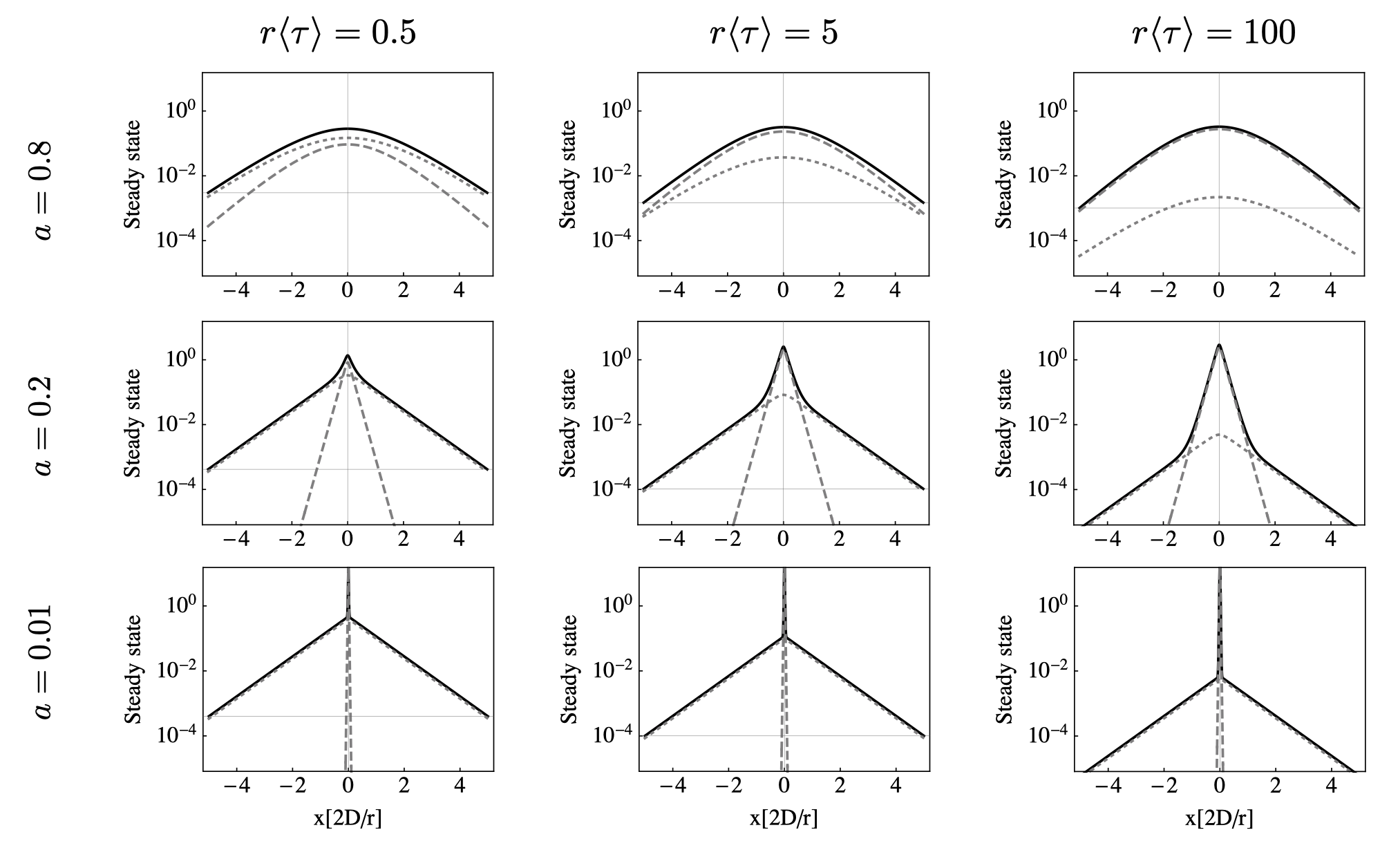}
    \caption{Steady states for a Brownian particle for various choices of resetting strength $a$ and refractory durations $\langle \tau \rangle$.  { \color{black} The solid line shows the full steady state, while the dotted and dashed line show the exploration and refractory population respectively.} At large $\langle\tau\rangle$ the trajectories trapped in the refractory long phases dominate the steady state. As the resetting becomes weaker $a \to 1$, a Gaussian is approached for any $\langle \tau \rangle$. Parameters chosen are $D = r = 1$. Steady states are obtained by truncating the infinite product in the analytical expressions at finite order and inverting the Fourier transform. We should not that this may require a small re-normalization, which has not been included in the above curves. This only shifts the curves and will not change any qualitative features.  }
    \label{fig:pd}
\end{figure}

The steady state in Eq.\,(\ref{eq:stst}) has a natural interpretation as a statistical co-existence of two populations; particles in the exploration phase and particles in the refractory phase. The particles in the exploration phase contribute to the steady state 
\begin{equation}\label{eq:E}
    \hat p_\mathcal{E}(k) = \prod_{j=0}^\infty r\hat{\tilde{p}}_0(a^j k, r |0) ,
\end{equation}
while the particles in the refractory phase contribute
\begin{equation}\label{eq:R}
    \hat p_\mathcal{R}(k) = \prod_{j=1}^\infty r\hat{\tilde{p}}_0(a^j k, r |0) .
\end{equation}
In the steady state, the probability of finding a particle in the exploration phase is given by the fraction of time spent in this phase, and reads $\pi_\mathcal{E} = (1 + r \langle \tau \rangle)^{-1}$, while the probability of finding a particle in the refractory phase is $\pi_\mathcal{R} = r \langle \tau \rangle (1 + r \langle \tau \rangle)^{-1}$. In terms of these quantities, the steady state can be written
\begin{equation}\label{eq:form}
\hat p_r^* (k) = \pi_\mathcal{E}\hat p_\mathcal{E}(k)  + \pi_\mathcal{R} \hat p_\mathcal{R}(k).
\end{equation}
Since $\hat p_{\mathcal{E},\mathcal{R}}(0) = 1$, normalization of $\hat p_r^* (k)$ follows immediately from $\pi_\mathcal{E}+ \pi_\mathcal{R} = 1$. As $\langle \tau \rangle$ changes from $0$ to $\infty$, the steady state transitions from $p_\mathcal{E}(x)$ to $p_\mathcal{R}(x)$. 
Fig.\,(\ref{fig:pd}) shows several steady states for a diffusion process, varying both the mean refractory time and the resetting strength. {\color{black} The co-existence of the exploration and refractory subpopulations can clearly be seen in the steady state; a sharper peak corresponding to particles stuck in refractory periods (dashed line), and a wider tail corresponding to exploring particles (dotted line).} We see that as resetting becomes weaker, the non-equilibrium steady state seems to approach a Gaussian for any value of the mean refractory time. As the mean refractory time is increased, more weight is given to the population in the refractory phase $p_\mathcal{R}(x)$. In the statistics literature, models with multiple components such as in Eq.\,(\ref{eq:form}) as referred to as mixture models \cite{everitt2013finite}, and has many applications in various complex systems.  For further quantitative insight into the non-equilibrium steady state in Eq.\,(\ref{eq:form}) we analyse the two contributions $p_\mathcal{S}(x)$, $\mathcal{S} = \mathcal{E},\mathcal{R}$, and study the associated moments and cumulants. We show that universal results can be derived for a wide range of underlying processes.

{\color{black}
\subsection{Moments and cumulants}

While the solution presented above is exact and can be studied numerically or semi-analytically by truncating the infinite products at finite values and inverting the Fourier transforms, one can gain further insights into the non-equilibrium steady state by considering the moments and cumulants.

\textbf{Moments --- } First, we note that due to the splitting of the non-equilibrium steady state into two subpopulations in Eq.\,(\ref{eq:form}), the moments are similarly given as 
\begin{equation}\label{eq:split}
\langle x^m \rangle_* = \pi_\mathcal{E} \langle x^m \rangle_\mathcal{E}   + \pi_\mathcal{R} \langle x^m \rangle_\mathcal{R}   ,
\end{equation}
where $\langle x^m \rangle_{\mathcal{S}}$ are the moments associated with the exploration and refractory phases $\mathcal{S} = \mathcal{E},\mathcal{R}$. Using similar methods to those of section \ref{sec:time}, we can also find explicit expressions for the steady state moments. Considering again a diffusion process where
\begin{equation}\label{eq:levy}
    \hat{\tilde p}(k,s) = \frac{1}{s + D k^2}.
\end{equation}
This implies that the two subpopulations of the steady state can again be written in terms of the q-Pochhammer symbol $(a;q)_n$ as in section \ref{sec:time}. For the exploration population, we have
\begin{equation}
     \hat{p}_\mathcal{E}(k) =\prod_{j=0}^\infty  \frac{1}{1 + \frac{D}{r} a^{2 j } k^\beta} = \left( -\frac{D}{r}k^2; a^2 \right)^{-1}_\infty.
\end{equation}
By using the series expansion for the q-Pochhammer symbol, Eq.\,(\ref{eq:qpoc}), we can write the series as
\begin{equation}
     \hat{p}_\mathcal{E}(k) = \sum_{n=0}^\infty \frac{(-1)^n}{(a^2;a^2)_n}\frac{D^n}{r^n} k^{2 n  }\equiv \sum_{n=0}^\infty C_\mathcal{E}(n) k^{2n}.
\end{equation}
For the refractory population, we note that since $ \hat{p}_\mathcal{R}(k) = ({1+\frac{D}{r}k^2})\hat{p}_\mathcal{E}(k)$, we have
\begin{equation}
     \hat{p}_\mathcal{R}(k) = 1 + \sum_{n=1}^\infty \left\{ C_\mathcal{E}(n) + \frac{D}{r}C_\mathcal{E}(n-1)\right\}  k^{2n} \equiv 1 + \sum_{n=1}^\infty C_\mathcal{R}(n) k^{2n}.
\end{equation}
With this series representation we can easily access moments for the subpopulations as well as for the full non-equilibrium steady state through Eq.\,(\ref{eq:split}). This can be compared with the series representation
\begin{equation}
     \hat{p}_{\mathcal{S}}(k) = \sum_{m=0}^\infty \frac{(-i)^m}{m!} \langle x^m\rangle_{\mathcal{S}} \:k^m
\end{equation}
 to find the moments
\begin{equation}\label{eq:exactdiffusion}
    \langle x^{2n}\rangle_{\mathcal{S}}  = (-1)^{n} {(2n)}!C_{\mathcal{S}}(n,2) =     \left\{
	\begin{array}{ll}
		\frac{{(2n)}!}{(a^2;a^2)_n}\frac{D^n}{r^n}    & \mbox{, if } \mathcal{S} =  \mathcal{E}, \\
		  \frac{{(2n)}!}{(a^2;a^2)_n}\frac{D^n}{r^n}  a^{2n}    & \mbox{, if } \mathcal{S} =  \mathcal{R},
	\end{array}
\right.
\end{equation}
which is valid for all orders. The full non-equilibrium steady state moments are then easily obtained from Eq.\,(\ref{eq:split}).

\textbf{Cumulants --- } While the above approach has the strength that is allows a closed-form expression for moments of all orders, it may be hard to generalize beyond diffusion or drift-diffusion processes. Furthermore, cumulants will not satisfy the same simple splitting rule Eq. (\ref{eq:split}) as the moments do, so the cumulants deserve some additional attention.  To better understand the cumulants, we consider their generating functions. Define the cumulant generating function of either of the two subpopulations $\mathcal{S} = \mathcal{E},\mathcal{R}$ as 
\begin{equation}
    \mathcal{G}_\mathcal{S}(k) = \log \hat p_\mathcal{S}(k),
\end{equation}
and let 
\begin{equation}
    \mathcal{C}_0(k) = \log ( r\hat {\tilde p}_0(k,r))
\end{equation}denote the cumulant generating function for the resetting problem with $a = \langle \tau \rangle =0$. From Eq.\,(\ref{eq:E}) and (\ref{eq:R}) we then immediately have
\begin{equation}
     \mathcal{G}_\mathcal{S}(k) = \sum_{j= j_0(\mathcal{S})}^\infty \mathcal{C}_0(a^j k),
\end{equation}
where the starting index of the sum depends on the population label as $j_0(\mathcal{E})=0$ and $j_0(\mathcal{R})=1$. The cumulants are then obtained by the application of $ \frac{\partial^m}{\partial(-i k)^m}$ and the subsequent limit $k\to 0$ of the corresponding generating function. Taking derivatives and using the chain rule, we have
\begin{equation}
    \kappa_m^{(\mathcal{S})} =  \left\{
	\begin{array}{ll}
		\kappa_m^{(0)}   \frac{1}{1-a^m}  & \mbox{, if } \mathcal{S} =  \mathcal{E}, \\
		 \kappa_m^{(0)}   \frac{a^m}{1-a^m}   & \mbox{, if } \mathcal{S} =  \mathcal{R},
	\end{array}
\right.  
\end{equation}
where $\kappa_m^{(0)}$ is the $m$'th cumulant for the $a = \langle \tau \rangle =0$ process. This immediately implies the relation 
\begin{equation}
    \frac{\kappa_m^{(\mathcal{R})} }{\kappa_m^{(\mathcal{E})} } = a^m,
\end{equation}
which holds independently of the underlying process and is a simple consequence of the fact that the $\mathcal{R}$-population can be thought of as the $\mathcal{E}$-population with rescaled coordinates $x\to a x$. Since $a\leq 1$, this also shows that $\kappa_m^{(\mathcal{R})}  \leq \kappa_m^{(\mathcal{E})} $ at all orders $m$. 

For the full non-equilibrium steady state, the cumulant generating function $\mathcal{G}(k) = \log \hat p_r^*(k)$ can also be obtained. From the general form of the non-equilibrium steady state, given in Eq.\,(\ref{eq:stst}), we have
\begin{equation}
  e^{\mathcal{G}(k)} =  \left[\frac{e^{  \mathcal{C}_0( k)} + r \langle \tau \rangle }{ 1+ r \langle \tau \rangle}   \right]e^{\sum_{j=1}^\infty  \mathcal{C}_0(a^j k)}  .
\end{equation}
We proceed to taking a logarithm and rearranging, which results in
\begin{equation}
  \mathcal{G}(k) = \log  \left[\frac{e^{  \mathcal{C}_0( k)} + r \langle \tau \rangle }{ 1+ r \langle \tau \rangle}   \right]  + \mathcal{G}_\mathcal{R}(k) {\color{black} =  \log  \left[\frac{e^{  \mathcal{C}_0( k)} + r \langle \tau \rangle }{ 1+ r \langle \tau \rangle}   \right]  + \sum_{j=1}^\infty \mathcal{C}_0(a^j k)}.
\end{equation}
{\color{black}We observe that the cumulant generating function in this way is entirely expressed in terms of the cumulant generating function $\mathcal{C}_0(k)$ for the same resetting problem with $a=0$ and no refractory periods. When calculating the cumulants from this expression by differentiation, this implies that the dependence on the mean refractory time and resetting strength is made explicit. }  For simplicity, we consider symmetric processes where $\partial_k^{2m+1}\mathcal{C}_0(k)|_{k=0}=0$. The first two non-zero cumulants are then given by
\begin{align}
   \sigma^2(x) &= \kappa_2 = \sigma_0^2 \left( \frac{  1}{1 + r \langle \tau \rangle} +  \frac{a^2}{1-a^2}\right),\\
   \kappa_4(x) &=  \sigma_0^4 \frac{ 3 r \langle \tau \rangle  }{[1 + r \langle \tau \rangle]^2} 
   + \kappa_4^{(0)} \left(  \frac{  1}{1 + r \langle \tau \rangle}  + \frac{a^4}{1-a^4} \right),
\end{align}
where $\sigma_0$ and $\kappa_4^{(0)}$ are the standard deviation and fourth cumulant of the problem without refractory periods and with $a =0$. We see that as the mean refractory duration grows, the variance of the steady state decreases. This is because particles on average spend a larger fraction of time trapped in a refractory phase, and has less time to freely explore and widen the steady state. 

The kurtosis can be calculated from the fourth cumulant as 
\begin{align}
    \mathbb{K}(a,\langle \tau\rangle) = \frac{\kappa_4(x)}{\sigma^4(x)} + 3.
\end{align}
This measures non-normality in the steady state, indicated by deviations from $\mathbb{K} = 3$. From the above, we see that 
\begin{align}
  \lim_{a\to 1}  \mathbb{K}(a,\langle \tau\rangle) = 3 +   \lim_{a\to 1}  \frac{\kappa_4^{(0)}(x)}{\sigma_0^4(x)} \frac{(1-a^2)^2}{1-a^4} = 3,
\end{align}
indicating that in the weak resetting limit the steady state approaches a Gaussian for any statistics of refractory times. This transition was first discovered in Ref. \cite{tal2022diffusion} for $\langle \tau \rangle = 0$ in the case of diffusion with or without a drift.

Furthermore, the kurtosis is non-monotonic as a function of mean refractory time for $0 < a < 1$. Using the above, one can readily verify that 
\begin{equation}
  \partial_{\langle \tau \rangle} \mathbb{K}(a,\langle \tau\rangle)  \sim \frac{1-a^2 r \langle \tau \rangle }{\left(a^2 r \langle \tau \rangle +1\right)^3} ,
\end{equation}
with a maximum at
\begin{equation}
    \langle \tau \rangle_* = \frac{1}{ra^2}.
\end{equation}
Hence, for fixed resetting strength, the kurtosis takes its maximum value for a specific choice of mean refractory time. We emphasize that this optimum is universal, in the sense that it holds without specifying the underlying system except symmetry and homogeneity. We also note that this effect is only seen for partial resetting, since $\langle \tau \rangle_*$ diverges for conventional resetting $a=0$.

}

\section{Discussion}\label{sec:concl}
The effect of refractory periods on partial stochastic resetting has been studied analytically. The Fourier-Laplace transform of the exact time-dependent propagator is obtained using the first renewal equation. Our analysis is general, without assumption regarding the reset-free system. For the case of diffusive processes, we have derived an exact expression for time-dependent moments of all orders, as well as their steady state values.

From the propagator the steady state is extracted, and we characterize its behaviour as a function of refractory duration and resetting strength. {\color{black} The steady state naturally splits into two subpopulations; in the stationary regime, a fraction of particles will reside in the exploration phase, while the remaining particles reside in the refractory phase.} The cumulant generating function for the steady state was derived, and expressed in terms of the associated complete (i.e. non-partial) resetting problem without refractory times, making the dependence on resetting strength and refractory times explicit. We show that in the limit of weak resetting, the steady state transition to a Gaussian for any choice of refractory times. We also showed that for symmetric processes there exists a universal value of the mean refractory time that maximizes the kurtosis of the steady state, that is determined only by the resetting rate and resetting strength, and is independent of the particular system under study. 

There are many potential outlooks related to partial resetting. In particular, it would be interesting to explore the relaxation towards the steady state in order to see if partial resetting shares some of the dynamical anomalies that are observed for conventional resetting \cite{evans2020review}. {\color{black} Furthermore, partial resetting under non-Poissonian inter-reset times would also be interesting to consider in the future.} 

\begin{acknowledgements}
The authors acknowledge support by the Deutsche Forschungsgemeinschaft (DFG) within the project LO 418/29-1. 
\end{acknowledgements}

\appendix
\section{Characteristic series for the propagator}\label{sec:moment_time}
To derive Eq.\,(\ref{eq:char}), we proceed by studying the two contributions to the propagator, $ \mathcal{P}_\mathcal{S} (x,s|x_0)$, $\mathcal{S}=\mathcal{E},\mathcal{R}$, independently. Using the series expansion of the q-Pochhammer symbol, Eq.\,(\ref{eq:qpoc}), one can easily show by rearranging the sum that 
\begin{equation}\label{eq:ap1}
     \mathcal{P}_\mathcal{E} (k,s|x_0) = \sum_{\ell=0}^\infty (-1)^\ell \left(  \sum_{n=0}^\infty {\ell + n \brack \ell}_{a^2} \frac{[r \tilde W(s)]^n}{(s+r)^{n+\ell +1}} d^\ell  \right) k^{2\ell}.
\end{equation}
The second term $ \mathcal{P}_\mathcal{R} (k,s|x_0) $ requires a bit more attention. We again start by using the series in Eq.\,(\ref{eq:qpoc}), which leads to 
\begin{align}
     \mathcal{P}_\mathcal{R} (k,s|x_0)  &=  r \frac{1-\tilde W(s)}{s}\sum_{n=1}^\infty\sum_{\ell = 0}^\infty  (-1)^\ell {\ell + n \brack \ell}_{a^2}\frac{[r \tilde W(s)]^{n-1}}{(s+r)^{n+\ell}}\left(1 + \frac{D k^2}{s+r}\right)     D^\ell k^{2\ell} \\
     &=  r \frac{1-\tilde W(s)}{s}\sum_{n=1}^\infty\sum_{\ell = 0}^\infty  (-1)^\ell {\ell + n \brack \ell}_{a^2}\frac{[r \tilde W(s)]^{n-1}}{(s+r)^{n+\ell}}    D^\ell k^{2\ell} \nonumber\\
    &+  r \frac{1-\tilde W(s)}{s}\sum_{n=1}^\infty\sum_{\ell = 0}^\infty (-1)^\ell  {\ell + n \brack \ell}_{a^2}\frac{[r \tilde W(s)]^{n-1}}{(s+r)^{n+\ell+1}}    D^{\ell+1} k^{2(\ell+1)}. 
\end{align}
Defining $m = \ell+1$ and rearranging gives 
\begin{align}
    \mathcal{P}_\mathcal{R} (k,s|x_0)   &=  \sum_{\ell = 0}^\infty  (-1)^\ell \left(  \sum_{n=1}^\infty {\ell + n \brack \ell}_{a^2} r \frac{1-\tilde W(s)}{s}  \frac{[r \tilde W(s)]^{n-1}}{(s+r)^{n+\ell}}    D^\ell \right)  k^{2\ell} \\
    &-  \sum_{m = 1}^\infty (-1)^{m}  \left(\sum_{n=1}^\infty {{m-1} + n \brack {m-1}}_{a^2}r \frac{1-\tilde W(s)}{s}  \frac{[r \tilde W(s)]^{n-1}}{(s+r)^{n+m}}    D^{m} \right) k^{2m}. 
\end{align}
Writing out the $\ell=0$ term and using the identity

\begin{equation}
    {{\ell} + n \brack {\ell}}_{a^2} - {{\ell-1} + n \brack {\ell-1}}_{a^2} = {{\ell} + n \brack {\ell}}_{a^2} \frac{ a^{2\ell} (1-a^{2n}) }{1 - a^{2(\ell+n)}}  
\end{equation}
allows us to combine the two sums, resulting in 
\begin{align}
   \mathcal{P}_\mathcal{R} (k,s|x_0) &=  \frac{1}{s} - \frac{1}{r + s - r \tilde W(s)} \nonumber \\
   &+ \sum_{\ell = 1}^\infty  (-1)^\ell \left(  \sum_{n=1}^\infty {{\ell} + n \brack {\ell}}_{a^2} \frac{ a^{2\ell} (1-a^{2n}) }{1 - a^{2(\ell+n)}}    r \frac{1-\tilde W(s)}{s}  \frac{[r \tilde W(s)]^{n-1}}{(s+r)^{n+\ell}}    D^\ell \right)  k^{2\ell} .
\end{align}
Combining with Eq.\,(\ref{eq:ap1}) results in the characteristic series Eq.\,(\ref{eq:char}) as desired.

\section{Derivation of the steady state}\label{sec:nesscalc}

To derive the steady state, Eq.\,(\ref{eq:stst}), we start with the propagator in the form
\begin{align}
        \hat{\tilde p}_r(k,s|x_0) &= \mathcal{P}_\mathcal{E} (k,s|x_0) +  \mathcal{P}_\mathcal{R} (k,s|x_0) \nonumber \\
        &=\sum_{n=0}^\infty \left[\frac{r \tilde W (s)}{s+r}\right]^n \frac{1}{s+r} e^{- i a^n k x_0}\prod_{j=0}^n (s+r)\hat{\tilde{p}}_0(a^j k, s+r |0) \nonumber \\
        & + r \frac{1-\tilde W(s)}{s}\sum_{n=1}^\infty \left[\frac{r \tilde W (s)}{s+r}\right]^{n-1}  \frac{1}{s+r} e^{- i a^n k x_0}\prod_{j=1}^n (s+r)\hat{\tilde{p}}_0(a^j k, s+r |0).
\end{align}
For simplicity, we consider $x_0 =0$. We first consider the contribution from the first term $\mathcal{P}_\mathcal{E} (k,s|0)$. We want to perform the limit
\begin{align}
        \lim_{s\to 0} s \mathcal{P}_\mathcal{E} (k,s|0) =  \lim_{s\to 0} s \sum_{n=0}^\infty \left[\frac{r \tilde W (s)}{s+r}\right]^n \frac{1}{s+r} \prod_{j=0}^n (s+r)\hat{\tilde{p}}_0(a^j k, s+r |0) .
\end{align}
To proceed, we note that the only way in which poles in $s$ can appear is due to the infinite sum, as all other terms approach constants as $s\to 0$. We can write
\begin{align}
        \lim_{s\to 0} s \mathcal{P}_\mathcal{E} (k,s|0) =  \lim_{s\to 0} s \sum_{n=0}^\infty \left[\frac{r \tilde W (s)}{s+r}\right]^n \frac{1}{r}\prod_{j=0}^n r \hat{\tilde{p}}_0(a^j k, r |0) .
\end{align}
Next we observe that the product $\prod_{j=0}^n r \hat{\tilde{p}}_0(a^j k, r |0) $ converges to a constant as $n\to \infty$ due to the fact that $a<1$. {\color{black} This is because $r \hat{\tilde{p}}_0(a^j k, r |0)  \to 1$ for large values of $j$.} Based on this, we assume next that for some large but finite cut-off value $n_*$ we can decompose the sum as
\begin{align}
        \lim_{s\to 0} s \mathcal{P}_\mathcal{E} (k,s|0) \approx  \lim_{s\to 0} s \sum_{n=0}^{n_*} \left[\frac{r \tilde W (s)}{s+r}\right]^n \frac{1}{r} \prod_{j=0}^n r \hat{\tilde{p}}_0(a^j k, r |0) 
        +\lim_{s\to 0} s \sum_{n=n_*}^\infty \left[\frac{r \tilde W (s)}{s+r}\right]^n \frac{1}{r} \prod_{j=0}^\infty r \hat{\tilde{p}}_0(a^j k, r |0) ,
\end{align}
where we in the second sum replaced the finite product with an infinite one. This approximation can be made arbitrarily accurate by choosing a sufficiently large value of $n_*$, and in the limit $n_*\to \infty$ the expression is exact. Next, observe that the finite sum in the first term is simply a polynomial $q_0 + q_1 s +...+ q_{n_*} s^{n_*}$ of finite order, since $\frac{r \tilde W (s)}{s+r} \approx 1 - (1+ r \langle \tau \rangle) s / r +...$ at small $s$. Upon multiplication with $s$ and letting $s\to 0$ this term vanishes. Hence
\begin{align}\label{eq:a1}
        \lim_{s\to 0} s \mathcal{P}_\mathcal{E} (k,s|0) \approx \lim_{s\to 0} s \sum_{n=n_*}^\infty \left[\frac{r \tilde W (s)}{s+r}\right]^n \frac{1}{r} \prod_{j=0}^\infty r \hat{\tilde{p}}_0(a^j k, r |0) .
\end{align}
Performing the geometric series and using the expansion $ \tilde W(s) = 1 - s \langle \tau \rangle + ...$ we arrive at

\begin{align}
        \lim_{s\to 0} s \mathcal{P}_\mathcal{E} (k,s|0) \approx \lim_{s\to 0}  \left( \frac{r}{1 + r \langle \tau \rangle} + \frac{s}{1 + r \langle \tau \rangle} - n_* s + \mathcal{O}(s^2) \right)\frac{1}{r} \prod_{j=0}^\infty r \hat{\tilde{p}}_0(a^j k, r |0) .
\end{align}
Taking the limit $s\to 0$ faster than the limit $n_* \to \infty$ we find the steady state value 
\begin{align}
        \lim_{s\to 0} s \mathcal{P}_\mathcal{E} (k,s|0) = \frac{1}{1 + r \langle \tau \rangle} \prod_{j=0}^\infty r \hat{\tilde{p}}_0(a^j k, r |0) .
\end{align}
We note that here we used the expansion $ \tilde W(s) = 1 - s \langle \tau \rangle + ...$, which assumes finite mean refractory times. The case of infinite $\langle \tau \rangle$ is treated separately in appendix \ref{sec:infmean}. 

Proceeding similarly with the second  term, we find 
\begin{equation}
   \lim_{s\to 0} s \mathcal{P}_\mathcal{R} (k,s|0) =  \frac{r \langle \tau \rangle}{1 + r \langle \tau \rangle} \prod_{j=1}^\infty r\hat{\tilde{p}}_0(a^j k, r |0).
\end{equation}

Combining these results, one arrives at the steady state in Eq.\,(\ref{eq:stst}).

\section{The case of infinite mean refractory time}\label{sec:infmean}
In deriving the non-equilibrium steady state in section \ref{sec:ness}, we expanded $\tilde W(s)$ to reveal the dependence on the mean refractory time. When the mean refractory time diverges, the steady state must be examined more carefully. 
Indeed, consider the case of a fixed refractory time $W(\tau) = \delta(\tau-\tau_0)$. The non-equilibrium steady state in Eq.\,(\ref{eq:stst}) is then reached for times $t \gg \tau_0$, since at least one (and in principle infinitely many) resetting events must take place for the stationary regime to be reached. However, since in the $\tau_0\to \infty$ limit the system is trapped indefinitely already in its first refractory period, the non-equilibrium steady state in Eq.\,(\ref{eq:stst}) is never reached. Considering this limit explicitly, we start with the full propagator
\begin{align}
        \hat{\tilde p}_r(k,s|x_0) &=\sum_{n=0}^\infty [r \exp(- s \tau_0)]^n  e^{- i a^n k x_0}\prod_{j=0}^n \hat{\tilde{p}}_0(a^j k, s+r |0) \nonumber \\
        & + r \frac{1-\exp(- s \tau_0)}{s}\sum_{n=1}^\infty [r \exp(- s \tau_0)]^{n-1} e^{- i a^n k x_0}\prod_{j=1}^n \hat{\tilde{p}}_0(a^j k, s+r |0).
\end{align}
When $\tau_0\to \infty$ only the first terms of the sums contribute, resulting in
\begin{align}
        \hat{\tilde p}_r(k,s|x_0) &=  e^{- i k x_0}  \hat{\tilde{p}}_0( k, s+r |0) +  \frac{r}{s}  e^{- i a k x_0}  \hat{\tilde{p}}_0(a k, s+r |0).
\end{align}
By the final value theorem, Eq.\,(\ref{eq:finalval}), the steady state becomes 
\begin{align}
        \hat{ p}_r^*(k|x_0) &=   r  e^{- i a k x_0}  \hat{\tilde{p}}_0(a k, r |0).
\end{align}
Inverting the Fourier transform, we arrive at 
\begin{align}\label{eq:infstst}
         p_r^*(x|x_0) =  \int_0^\infty  dt \psi(t) \frac{\rho_0(x/a,t|x_0)}{a} ,
\end{align}
with $\psi(t) = r e^{-rt}$. This steady state has a rather simple interpretation; the probability density at the time just before the first reset is simply $\int_0^\infty  dt \psi(t) \rho_0(x,t|x_0)$, after which a reset $x\to a x$ takes place. The process is then stuck at this location since the refractory time is infinite. By basic transformation rules of probability densities, this first resetting position of the particle has a density that is given exactly by Eq.\,(\ref{eq:infstst}).

It is worth emphasising that this steady state still depends on the initial position $x_0$, in contrast to the case of conventional resetting. Indeed, taking the $a\to 0$ limit, we see that $ \rho_0(x/a,t|x_0)/a$ vanishes for large values of its argument, while it diverges at $x=0$. Hence,
\begin{equation}
   \lim_{a\to 0}  p_r^*(x|x_0)  = \delta(x),
\end{equation}
which is independent of $x_0$ as found in Ref. \cite{evans2018effects}.


%

\end{document}